# biastest: Testing parameter equality across different models in Stata

Hasraddin Guliyev[1]

[1] Azerbaijan State University of Economic, Baku, AZ1007, Azerbaijan.
E-mail: hasradding@unec.edu.az

**Abstract**
The biastest command in Stata is a powerful and user-friendly tool designed to compare the coefficients of different regression models, enabling researchers to assess the robustness and consistency of their empirical findings. This command is particularly valuable for evaluating alternative modeling approaches, such as ordinary least squares versus robust regression, robust regression versus median regression, quantile regression across different quantiles, and fixed effects versus random effects models in panel data analysis. By providing both variable-specific and joint tests, biastest command offers a comprehensive framework for detecting bias or significant differences in model estimates, ensuring that researchers can make informed decisions about model selection and interpretation.

**Keywords:** *bias test, model selection, model interpretation*

## 1. Introduction

In empirical research, it is often necessary to compare the results of different regression models to assess the robustness of findings. For example, researchers may want to compare coefficients from an ordinary least squares (OLS) model with those from a robust regression model to determine whether outliers or influential observations are biasing the results (Rousseeuw & Leroy, 2003). Similarly, comparisons between robust regression and quantile regression, or between quantile regressions at different quantiles, can provide insights into how relationships between variables vary across the distribution of the dependent variable (Koenker & Bassett, 1978).

While Stata provides tools for estimating various models, there is no built-in command to directly test the equality of parameters across models. This limitation makes it challenging for researchers to systematically compare coefficients from different modeling approaches.

This paper introduces biastest command, a Stata program designed to facilitate comparisons between regression models. The new command computes differences in coefficients across two models, tests their statistical significance, and provides both individual and joint tests for parameter equality. Its flexibility allows users to specify any two regression models and their respective options, enabling comparisons such as OLS versus robust regression, robust regression versus median regression, quantile regressions at different quantiles, and fixed effects versus random effects models in panel data analysis. This versatility makes biastest a valuable tool for researchers seeking to assess the robustness and distributional implications of their empirical results.

The paper is structured as follows: Section 2 discusses the methodology behind the biastest command. It begins with a review of the statistical framework for comparing coefficients across different regression models and then introduces the individual and joint tests for parameter equality. Section 3 details the syntax and usage of the biastest command, highlighting its flexibility in accommodating various regression models and options. Section 4 provides several practical examples demonstrating the application of the biastest command in different contexts. Finally, Section 5 concludes the paper, summarizing contributions of biastest command and its potential to enhance empirical research.

## 2. Methodology

The biastest command is designed to statistically compare regression coefficients from two models, a procedure commonly employed in econometrics and social sciences to assess the robustness and consistency of model estimates.

Let $\hat{\beta}^1$ and $\hat{\beta}^2$ denote the coefficient vectors from Model 1 and Model 2, respectively. The test evaluates the null hypothesis:

$$H_0: \hat{\beta}^1 = \hat{\beta}^2 \quad (Eq. 1)$$

This hypothesis is tested through both individual parameter comparisons via t-test and a joint test of equality via a Wald-type chi-squared test, as discussed in Wooldridge (2010) and Greene (2018). For each independent variable j, the command computes the difference between the coefficients obtained from the two models:

$$\Delta\hat{\beta}_j = \hat{\beta}_j^1 - \hat{\beta}_j^2 \ (Eq. 2)$$

where $\hat{\beta}_j^1$ and $\hat{\beta}_j^2$ are the estimated coefficients of the j-th independent variable in Model 1 and Model 2, respectively. The standard error of this difference is calculated as:

$$SE(\Delta\hat{\beta}_j) = \sqrt{SE(\hat{\beta}_j^1)^2 + SE(\hat{\beta}_j^2)^2} \quad (Eq. 3)$$

This standard error assumes that the estimates are independent, and their variances are additive. The command uses the delta method to compute these standard errors, which is a conventional analytical approach when comparing independent model estimates (Clogg et al., 1995; Paternoster et al., 1998). The corresponding t-statistic is then:

$$t_j = \frac{\Delta\hat{\beta}_j}{SE(\Delta\hat{\beta}_j)} \quad (Eq. 4)$$

and the associated p-value is derived using the t-distribution with degrees of freedom equal to $N - k$, where $N$ is the sample size and $k$ is the number of independent variables. Both models must include the exact same dependent and independent variables in the same order. This is enforced through the command syntax. However, users should avoid using differing functional forms (e.g., interaction terms or factor variable expansions) between models, as this would violate the comparability assumption.

The models are assumed to be non-nested, and the command is not suitable for nested model comparisons. This design ensures that the coefficient differences are meaningful and interpretable. In addition to individual t-tests, the command performs a joint Wald chi-squared test for the equality of all coefficients (excluding the constant):

$$\chi^2 = (\hat{\beta}_1 - \hat{\beta}_2)'(\hat{V}_1 - \hat{V}_2)^{-1}(\hat{\beta}_1 - \hat{\beta}_2) \quad (Eq. 5)$$

where $V_1$ and $V_2$ are the variance-covariance matrices of the coefficient estimates from Model 1 and Model 2 (Greene, 2018).

Unlike existing informal approaches—such as visually comparing confidence interval overlaps or manually coding Wald tests—the biastest command offers a structured and automated solution. Notably, it allows users the flexibility to apply either bootstrap or robust confidence intervals when interpreting coefficient differences, which enhances its adaptability to various empirical contexts. Users are strongly advised to verify that models are correctly specified and estimated on well-conditioned data. The command does not detect multicollinearity, small-sample bias, or model misspecification, which may affect the reliability of test results. Warnings are issued for small sample sizes ($N - k < 30$) and non-positive-definite variance-covariance matrices, especially during the joint test. If the test matrix is not invertible or stable, users are advised to consider the sigmaless option.

## 3. The biastest command

The biastest command compares the coefficients of two regression models and tests the equality of their parameters. It provides both individual and joint tests for parameter equality, making it a useful tool for assessing the robustness and consistency of regression results across different modeling approaches.

### 3.1 Syntax

The syntax of biastest is as follows:

biastest depvar indepvars [if] [in], m1(string) [m1ops(string)] m2(string) [m2ops(string)] [SIGMALESS]

### 3.2 Options

m1(string) - Specifies the first regression model to be estimated. This is a required option. Example: m1(regress)

m1ops(string) - Specifies additional options to be passed to the first model. This is optional. Example: m1ops(robust)

m2(string) - Specifies the second regression model to be estimated. This is a required option. Example: m2(rreg)

m2ops(string) - Specifies additional options to be passed to the second model. This is optional. Example: m2ops(nolog)

**SIGMALESS** *(optional)* - If specified, applies a scaling adjustment to the smaller model's variance-covariance matrix based on the relative RMSE of the two models. This is useful when the standard errors differ systematically between models.

### 3.3 Stored results

The biastest command stores the following results in e():

**Matrices**

e(b_m1) – Coefficient vector from the first model

e(V_m1) – Variance-covariance matrix from the first model

e(b_m2) – Coefficient vector from the second model

e(V_m2) – Variance-covariance matrix from the second model

e(t_stats) – t-statistics for individual parameter tests

e(p_values) – p-values for individual parameter tests

e(diffs) – Differences in coefficients between models

e(ses) – Standard errors of coefficient differences

**Scalars**

e(N) – Number of observations used

e(chi2) – Chi-squared statistic for the joint test

e(p) – p-value for the joint chi-squared test

e(df) – Degrees of freedom for the joint test

e(s2_1) – Estimated variance from Model 1 *(only if SIGMALESS is used)*

e(s2_2) – Estimated variance from Model 2 *(only if SIGMALESS is used)*

## 4. Examples

### 4.1 Comparison of ordinal least squares and robust regression

For our first example, we will use the crime dataset. This dataset featured Statistical Methods for Social Sciences, Third Edition, by Alan Agresti and Barbara Finlay (1997). The variables are state id (sid), state name (state), violent crimes per 100,000 people (crime), the percent of the population living in metropolitan areas (pctmetro), percent of population with a high school education or above (pcths), and percent of population living under poverty line (poverty). The dataset contains 51 observations, one for each U.S. state. In this analysis, we will use the biastest command to compare the results of OLS (reg) with those of robust regression (rreg).

### Example 1. Compare the results of OLS regression and robust regression

```
. use https://stats.idre.ucla.edu/stat/stata/dae/crime
(crime data from agresti & finlay - 1997)

. . biastest crime pctmetro pcths poverty , m1(reg) m2(rreg) m2ops(nolog)
Dependent variable: crime
Independent variables: pctmetro pcths poverty

Model 1 Estimation Results

      Source |       SS           df       MS      Number of obs   =        51
-------------+----------------------------------   F(3, 47)        =     28.20
       Model |  6253726.47         3  2084575.49   Prob > F        =    0.0000
    Residual |  3474748.27        47  73930.8143   R-squared       =    0.6428
-------------+----------------------------------   Adj R-squared   =    0.6200
       Total |  9728474.75        50  194569.495   Root MSE        =     271.9

------------------------------------------------------------------------------
       crime | Coefficient  Std. err.      t    P>|t|     [95% conf. interval]
-------------+----------------------------------------------------------------
    pctmetro |   11.92801    1.75924     6.78   0.000     8.388875    15.46714
       pcths |   26.86559   10.31791     2.60   0.012     6.108642    47.62255
     poverty |   76.86437   12.60933     6.10   0.000     51.49767    102.2311
       _cons |  -3334.774   946.0239    -3.53   0.001    -5237.928   -1431.619
------------------------------------------------------------------------------

Model 2 Estimation Results

Robust regression                               Number of obs     =         50
                                                F(  3,        46) =      22.04
                                                Prob > F          =     0.0000

------------------------------------------------------------------------------
       crime | Coefficient  Std. err.      t    P>|t|     [95% conf. interval]
-------------+----------------------------------------------------------------
    pctmetro |   9.547002   1.378778     6.92   0.000     6.771666    12.32234
       pcths |   4.582318   8.272346     0.55   0.582    -12.06905    21.23369
     poverty |   40.77088   10.99935     3.71   0.001     18.63035    62.91142
       _cons |  -1000.856   780.8014    -1.28   0.206    -2572.527    570.8159
------------------------------------------------------------------------------

Variable Bias Test:
-------------------------------------------------------------------------------
H0: The parameters are equal
Variable           Model 1      Model 2       Diff.        Std. Err.    t-stat        P>|t|
-------------------------------------------------------------------------------
 pctmetro          11.9280       9.5470       2.3810       1.0927       2.179        0.034
 pcths             26.8656       4.5823      22.2833       6.1666       3.614        0.001
 poverty           76.8644      40.7709      36.0935       6.1652       5.854        0.000
-------------------------------------------------------------------------------

Joint Bias Test:
H0: All parameters are equal
----------------------------------
 chi2(3)      =    46.680
 Prob > chi2  =     0.000
----------------------------------
```

The variable bias test and joint bias test are statistical tools used to compare the results of two regression models—in this case, OLS and robust regression—to determine if there are significant differences in their estimated coefficients. The variable bias test examines each independent variable individually, while the joint bias test evaluates whether all coefficients are jointly equal across the two models. These tests are particularly useful for assessing the robustness of regression results, especially in the presence of outliers or influential observations.

The variable bias test revealed notable differences in the coefficients for key predictors of violent crime. For urbanization (pctmetro), the difference between OLS and robust regression is statistically significant at 5% level (p-value = 0.034 < 0.05), suggesting that the effect of urbanization on crime is influenced by outliers. In addition, for education (pcths), the difference is significant at 5% level (p-value = 0.001 < 0.05). The OLS model estimated a stronger positive effect, while the robust regression model showed a weaker and statistically insignificant effect, indicating that the effect of education on crime may be sensitive to outliers. The most striking difference was for poverty (poverty), where the difference in coefficients was statistically significant at %5 level (p-value = 0.000 < 0.05). The OLS estimated a much larger effect of poverty on crime compared to the robust regression model, highlighting the sensitivity of this relationship to outliers.

The joint bias test further confirmed that the two models are not equivalent. The test statistic $\chi^2$ (3) = 46.680 and p-value = 0.000 < 0.05 led to the rejection of the null hypothesis at 5% significance level that all coefficients are equal across the two models. This indicates that the choice of regression method significantly affects the estimated relationships between independent variables and violent crime. The robust regression results, which are less sensitive to outliers, suggest that poverty and urbanization are the most significant predictors of crime, while education has a weaker and statistically insignificant effect. In conclusion, the bias tests highlight that robust regression offers more reliable estimates for variables, particularly in the presence of outliers or influential observations (Huber and Ronchetti, 2009; Rousseeuw and Leroy, 2003; Verardi and Croux, 2009).

**Example 2. Compare the results of robust regression and median regression**

```
. biastest crime pctmetro pcths , m1(rreg) m1ops(nolog) m2(qreg) m2ops(nolog)
Dependent variable: crime
Independent variables: pctmetro pcths

Model 1 Estimation Results

Robust regression                                 Number of obs     =         51
                                                  F(  2,        48) =      20.54
                                                  Prob > F          =     0.0000

------------------------------------------------------------------------------
       crime | Coefficient  Std. err.      t    P>|t|     [95% conf. interval]
-------------+----------------------------------------------------------------
    pctmetro |   8.623612   1.530345     5.64   0.000     5.546647    11.70058
       pcths |  -18.21422   6.008845    -3.03   0.004    -30.29582   -6.132631
       _cons |   1376.441    471.059     2.92   0.005     429.3134    2323.569
------------------------------------------------------------------------------

Model 2 Estimation Results

Median regression                                     Number of obs =         51
  Raw sum of deviations     7393 (about 515)
  Min sum of deviations 5190.675                      Pseudo R2     =     0.2979

------------------------------------------------------------------------------
       crime | Coefficient  Std. err.      t    P>|t|     [95% conf. interval]
-------------+----------------------------------------------------------------
    pctmetro |   9.532475   2.071695     4.60   0.000     5.367052     13.6979
       pcths |  -19.27213   8.134439    -2.37   0.022    -35.62752   -2.916747
       _cons |   1413.812   637.6934     2.22   0.031      131.643     2695.98
------------------------------------------------------------------------------

Variable Bias Test:
--------------------------------------------------------------------------------------------
H0: The parameters are equal
Variable           Model 2      Model 1        Diff.       Std. Err.     t-stat        P>|t|
--------------------------------------------------------------------------------------------
 pctmetro           9.5325       8.6236       0.9089        1.3964        0.651        0.518
 pcths            -19.2721     -18.2142      -1.0579        5.4830       -0.193        0.848
--------------------------------------------------------------------------------------------

Joint Bias Test:
H0: All parameters are equal
------------------------------------
 chi2(2)      =     0.462
 Prob > chi2  =     0.794
------------------------------------
```

The variable biastest compares the coefficients of each independent variable between robust regression and median regression to assess whether there are significant differences in their estimates. These regressions are alternative models that are particularly useful for addressing the influence of outliers or extreme observations. The test results imply that both urbanization (pctmetro) and education (pcths) have reliable and consistent effects on crime, regardless of whether robust regression or median regression is used. The joint bias test evaluates whether all coefficients in the two models are jointly equal. The test statistic $\chi^2$ (2) = 0.462 and p-value = 0.794 > 0.05 indicates that the null hypothesis of equal coefficients cannot be rejected at 5% significance level. This means that the coefficients from robust regression and median regression are not significantly different when considered jointly. The consistency between the two models underscores the robustness of the findings, suggesting that the choice of regression models does not significantly alter the estimated relationships between the predictors and crime.

**4.2 Comparison across quantiles**

Quantile regression has gained significant attention in recent years due to its ability to provide a more comprehensive understanding of the relationship between variables across different points of the conditional distribution of the dependent variable. Unlike ordinary least squares (OLS) regression, which focuses on the mean of the dependent variable, quantile regression allows researchers to analyze the impact of independent variables at various quantiles (e.g., 10th, 25th, 50th, 75th, 90th percentiles). This is particularly useful in fields such as economics, finance, and social sciences, where the effects of predictors may vary across the distribution (Koenker & Bassett, 1978; Koenker, 2005). One of the key advantages of quantile regression is its robustness to outliers and non-normal distributions, making it a valuable tool for analyzing data with heterogeneous patterns (Cade & Noon, 2003; Davino et al., 2014). For example, in income distribution studies, the impact of education or experience on income might differ significantly between the lower and upper quantiles. Quantile regression captures these nuances, providing deeper insights into the underlying data structure (Buchinsky, 1998; Machado & Mata, 2005).

However, as the use of quantile regression has increased, so has the importance of testing the equality of coefficients across different quantiles. This is crucial because it helps determine whether the relationships between variables are consistent throughout the distribution or if they vary significantly at different points (Hao & Naiman, 2007; Angrist & Pischke, 2009). For instance, if the coefficient of an independent variable at the 10th quantile is significantly different from that at the 90th quantile, it suggests that the variable's effect is not uniform across the distribution. Such findings can have important implications for policy-making and theoretical models, as they reveal heterogeneous effects that may be masked by traditional regression methods (Chernozhukov & Hansen, 2005; Powell, 2016).

We explore the relationship between household income and food expenditure using the data from Engel (1857), as described in Koenker and Bassett (1978). We will use simultaneous quantile regression to compare this relationship across different quantiles. To test for bias or differences in the relationship across quantiles, we use the biastest command. This command compares the quantile regression models at the 25th percentile q(.25) and the 75th percentile q(.75), with 100 replications r(100) to calculate bootstrap standart errors for each model.

## Example 3. Compare the results of different quantiles

```
. webuse engel1857
(European household budget survey)

. set seed 1234

. biastest foodexp income, m1(sqreg) m1ops(q(.25) r(100) nolog) m2(sqreg) m2ops(q(.75) r(100) no
Dependent variable: foodexp
Independent variables: income

Model 1 Estimation Results

Simultaneous quantile regression                    Number of obs =        235
  bootstrap(100) SEs                                .25 Pseudo R2 =     0.5540

------------------------------------------------------------------------------
             |              Bootstrap
     foodexp | Coefficient  std. err.      t    P>|t|     [95% conf. interval]
-------------+----------------------------------------------------------------
q25          |
      income |   .4741032   .0307464    15.42   0.000     .4135268    .5346796
       _cons |   .0954835   .0234381     4.07   0.000     .0493059    .1416611
------------------------------------------------------------------------------

Model 2 Estimation Results

Simultaneous quantile regression                    Number of obs =        235
  bootstrap(100) SEs                                .75 Pseudo R2 =     0.6966

------------------------------------------------------------------------------
             |              Bootstrap
     foodexp | Coefficient  std. err.      t    P>|t|     [95% conf. interval]
-------------+----------------------------------------------------------------
q75          |
      income |   .6440143   .0282025    22.84   0.000     .5884498    .6995787
       _cons |   .0623965   .0228063     2.74   0.007     .0174635    .1073294
------------------------------------------------------------------------------

Variable Bias Test:
-------------------------------------------------------------------------------------------
H0: The parameters are equal
Variable           Model 2      Model 1       Diff.        Std. Err.     t-stat       P>|t|
-------------------------------------------------------------------------------------------
 income             0.6440       0.4741       0.1699       0.0122       13.875        0.000
-------------------------------------------------------------------------------------------

Joint Bias Test:
H0: All parameters are equal
------------------------------------
 chi2(1)      =   192.518
 Prob > chi2  =     0.000
------------------------------------
```

The analysis explores the relationship between household income and food expenditure using quantile regression, as implemented in Stata with the biastest command. The results reveal significant heterogeneity in the income-food expenditure relationship across the distribution of food expenditure. At the 25th percentile, a one-unit increase in household income is associated with a 0.474-unit increase in food expenditure, with a pseudo $R^2$ of 0.554. In contrast, at the 75th percentile, the same one-unit increase in income corresponds to a 0.644-unit increase in food expenditure, with a higher pseudo $R^2$ of 0.697. These findings suggest that the income elasticity of food expenditure is stronger for higher-income households compared to lower-income households. This pattern indicates that as households move up the income distribution, a larger proportion of their income is allocated to food expenditure, reflecting differences in consumption behavior and priorities across income groups.

The biastest command was used to formally compare the coefficients across the 25th and 75th percentiles. The variable bias test indicates a statistically significant difference in the income coefficient at the 5% level between the two quantiles, with a difference of 0.1699 and p-value=0.000 < 0.05. Additionally, the jointly bias test calculate $\chi^2(1)$ = 192.518 ($t^2=\chi^2$; $13.875^2=192.518$) and p-value = 0.000 < 0.05 which rejects the null hypothesis that all parameters are equal across the two models, further confirming the heterogeneity in the relationship. These results underscore the importance of using quantile regression to capture varying effects across the distribution of the outcome variable.

### 4.3 Comparison of panel data models

The Hausman test is a widely used statistical tool in econometrics to determine whether a fixed effects model or a random effects model is more appropriate for panel data analysis. This test is crucial because it helps researchers select the correct model specification, ensuring that the estimates are both accurate and reliable. Choosing the wrong model can lead to biased and inconsistent estimates, which can significantly affect the validity of empirical findings (Wooldridge, 2010; Greene, 2018).

### Example 4. Compare the panel data models

```
. . webuse grunfeld

. biastest invest mvalue kstock, m1(xtreg) m1ops(fe) m2(xtreg) m2ops(re)
Dependent variable: invest
Independent variables: mvalue kstock

Model 1 Estimation Results

Fixed-effects (within) regression               Number of obs     =        200
Group variable: company                         Number of groups  =         10

R-squared:                                      Obs per group:
     Within  = 0.7668                                         min =         20
     Between = 0.8194                                         avg =       20.0
     Overall = 0.8060                                         max =         20

                                                F(2, 188)         =     309.01
corr(u_i, Xb) = -0.1517                         Prob > F          =     0.0000

------------------------------------------------------------------------------
      invest | Coefficient  Std. err.      t    P>|t|     [95% conf. interval]
-------------+----------------------------------------------------------------
      mvalue |   .1101238   .0118567     9.29   0.000     .0867345    .1335131
      kstock |   .3100653   .0173545    17.87   0.000     .2758308    .3442999
       _cons |  -58.74393   12.45369    -4.72   0.000    -83.31086     -34.177
-------------+----------------------------------------------------------------
     sigma_u |  85.732501
     sigma_e |  52.767964
         rho |  .72525012   (fraction of variance due to u_i)
------------------------------------------------------------------------------
F test that all u_i=0: F(9, 188) = 49.18                     Prob > F = 0.0000

Model 2 Estimation Results

Random-effects GLS regression                   Number of obs     =        200
Group variable: company                         Number of groups  =         10

R-squared:                                      Obs per group:
     Within  = 0.7668                                         min =         20
     Between = 0.8196                                         avg =       20.0
     Overall = 0.8061                                         max =         20

                                                Wald chi2(2)      =     657.67
corr(u_i, X) = 0 (assumed)                      Prob > chi2       =     0.0000

------------------------------------------------------------------------------
      invest | Coefficient  Std. err.      z    P>|z|     [95% conf. interval]
-------------+----------------------------------------------------------------
      mvalue |   .1097811   .0104927    10.46   0.000     .0892159    .1303464
      kstock |    .308113   .0171805    17.93   0.000     .2744399    .3417861
       _cons |  -57.83441   28.89893    -2.00   0.045    -114.4753   -1.193537
-------------+----------------------------------------------------------------
     sigma_u |   84.20095
     sigma_e |  52.767964
         rho |  .71800838   (fraction of variance due to u_i)
------------------------------------------------------------------------------

Variable Bias Test:
-------------------------------------------------------------------------------------------
H0: The parameters are equal
Variable            Model 1      Model 2       Diff.        Std. Err.    t-stat       P>|t|
-------------------------------------------------------------------------------------------
 mvalue              0.1101       0.1098       0.0003        0.0055       0.062       0.951
 kstock              0.3101       0.3081       0.0020        0.0025       0.796       0.427
-------------------------------------------------------------------------------------------

Joint Bias Test:
H0: All parameters are equal
------------------------------------
 chi2(2)      =     2.330
 Prob > chi2  =     0.312
------------------------------------
```

The Hausman test operates by comparing the coefficient estimates from the fixed effects and random effects models. Under the null hypothesis, the random effects model is preferred because it is more efficient, assuming that the unobserved individual effects are uncorrelated with the explanatory variables. If the null hypothesis is rejected, the fixed effects model is preferred, as it controls for potential correlation between the unobserved effects and the regressors, thereby providing consistent estimates (Hausman, 1978). While performing the Hausman (1978) test manually in Stata can be somewhat cumbersome, the biastest command simplifies the process significantly. Below, we demonstrate how to use the biastest command with the Grunfeld (1958) dataset to compare fixed effects and random effects models and determine the most suitable approach.
The variable bias test compares the coefficients of the fixed effects (FE) and random effects (RE) models to determine whether the differences between them are statistically significant. In this analysis, the results of the variable bias test show that for mvalue, the difference between the FE and RE coefficients is 0.0003, with t-stat=0.062 and p-value = 0.951 > 0.05. Similarly, for kstock, the difference is 0.0020, with t-stat=0.796 and p-value = 0.427> 0.05. These results indicate that the differences in coefficients for both variables are not statistically significant at % 5 level. This suggests that the FE and RE models produce estimates that are not meaningfully different for these variables. Furhermore, the joint bias test or Hausman (1978) test further supports this conclusion. The test statistics is $\chi^2(2) = 2.3304$, with p-value = 0.312 > 0.05 indicates that there is no statistically significant difference between the FE and RE models as a whole. We fail to reject the null hypothesis that the coefficients of the FE and RE models are equal at 5% significance level. This result implies that the random effects model is appropriate for Grunfeld model, as there is no evidence of significant correlation between the unobserved heterogeneity and the regressors.

## 5. Conclusions

The biastest command in Stata is a powerful and versatile tool for comparing the coefficients of different regression models, enabling researchers to assess the robustness and consistency of their findings. Its applications span a wide range of contexts, including comparisons between ordinary least squares (OLS) and robust regression, robust regression and median regression, quantile regression across different percentiles, and fixed effects versus random effects models in panel data analysis. These examples highlight the command's ability to provide statistical evidence for choosing between alternative modeling approaches, particularly in the presence of outliers, heterogeneous effects, or unobserved heterogeneity.

By simplifying complex statistical comparisons, the biastest command makes it accessible for researchers to test for bias or differences across models. It offers both variable-specific and joint tests, providing a comprehensive approach to model comparison. This ensures that researchers can make informed decisions about model selection and interpretation, enhancing the reliability of their empirical work.

Overall, the biastest command represents a significant contribution to Stata's toolkit for statistical analysis. It empowers researchers to conduct more nuanced and reliable analyses, particularly when comparing alternative modeling strategies.

**Supplementary information**
To install the software files as they existed at the time of publication of this article in Stata:
*ssc install biastest*
*help biastest*

**Acknowledgements**
The authors declare that this study has no financial support.

**Authors' Contribution**
H.G contributed to writing—review & editing, writing—original draft, software, resources, methodology, formal analysis, and conceptualization.

**Data Availability**
No datasets were generated or analyzed during the current study.

**Conflict of interest**
The authors declare no competing interests.